\documentclass[preprint,12pt,amsmath,amssymb,showkeys]{revtex4} 
\usepackage{amsmath}
\usepackage{latexsym}
\usepackage{amssymb}
\usepackage{graphics,epstopdf}
\usepackage{amsmath,amssymb,amsthm}
\theoremstyle{plain}

\usepackage{graphicx}
\usepackage{enumerate}
\usepackage[colorlinks=true, citecolor=blue, urlcolor=blue ]{hyperref}

\begin{document}
\title{On the  $\mathcal{P}\mathcal{T}$-symmetric parametric amplifier }

\author{Pinaki Patra}
\thanks{Corresponding author}
\email{monk.ju@gmail.com}
\affiliation{Department of Physics, Brahmananda Keshab Chandra College, Kolkata, India-700108}

\date{\today}

\begin{abstract}
Parametric amplifiers are an integral part of measurements involving the conversion of propagating quantum information to mechanical motion. General time-dependent  PT-symmetric parametric oscillators for unbroken parity and time reversal (PT) symmetry  regimes are studied theoretically. By constructing an explicit metric operator, we have transformed the non-Hermitian PT-symmetric system to an equivalent Hermitian Hamiltonian, which enables us to utilize the available mechanism of $\mathbb{L}^2$ space. \\
The time-dependent (TD) Schr\"{o}dinger equation for the system is solved with the Lewis-Riesenfeld (LR) phase space method. The eigenstates of the LR-invariant operator ($\hat{\mathcal{I}}$) is obtained after transforming $\hat{\mathcal{I}}$ to its diagonal symplectic equivalent form (group $Sp(2, \mathbb{R})$).  Both the dynamical and geometrical phase factors associated with the eigenstates of $\hat{\mathcal{I}}$ are explicitly written. \\
The experimental pheasibility of our result is outlined through the construction of Wigner quasiprobability distribution. Moreover, we have demostrated the time variation of the Wigner distribution for the system consisting of two spatially separated prepared  ground state of the TD-parametric amplifier.  With graphical illustration of time variation of Wigner distributions, we show that the phase-space entanglement remains intact even for time-dependent situation, no matter how far the particles goes, at least for the cat-state under consideration. \\
The exact expressions for the physically relevant qualities are obtained and illustrated for a  toy model. 
\end{abstract}
\keywords{Parametric amplifier; PT-symmetric quantum theory;  Wigner distribution; Lewis-Riesenfeld invariance; Coherent states; Ermakov Pinney equation; Entanglement}
 
 \maketitle
\section{Introduction}
A parametric  amplifier, in principle, can support noiseless amplification \cite{amplifier1}.
For instance, a resonant Josephson parametric amplifier  provides high gain, quantum-limited noise performance, tunable center frequency, and is simple to manufacture within present-day industrial capacity \cite{amplifier2,amplifier3,amplifier4}. A continuous range of nonclassical states of light can be generated from an optical parametric  amplifier by post-selection  on the idler mode of a parametric amplifier \cite{amplifier5}. Interestingly, a lossless beam-splitter  and an optical parametric  amplifier are equivalent to each other when the input and output signals are appropriately interchanged \cite{amplifier6}. Beam-splitter is an inevitable part of any optical experiment through interferometry  \cite{interferometer1,interferometer2}.  Thus the equivalence between a lossless beam-splitter and an optical parametric  amplifier made the parametric amplification a topical one \cite{amplifier1,amplifier2,amplifier3,amplifier4,amplifier5,amplifier6,amplifier7,amplifier8}. 
\\
On the other hand, the requirement of real spectrum for a viable measurement in quantum mechanics can be served by not only the self-adjoint Hamiltonians but also a wider class of complex-non-Hermitian-Hamiltonians admitting parity ($\hat{\mathcal{P}}$) and time-reversal ($\hat{\mathcal{T}}$) symmetry \cite{pt1,pt2}. After the initial proposal of Bender and Boettcher \cite{pt1}, a considerable amount of literature is flooding  into identifying properties of physical systems described by $\hat{\mathcal{P}}\hat{\mathcal{T}}$ -symmetric Hamiltonians.
Such Hamiltonians, although not Hermitian, are consistent to generate unitary time evolutions  with a preferentially selected inner
product ($\langle u\vert \hat{\eta}v\rangle$) from an appropriate metric operator $\hat{\eta}$ \cite{pt3,pt4,pt5,pt51,pt52}. For a consistent introduction to the axiomatic $\hat{\mathcal{P}}\hat{\mathcal{T}}$ -symmetric quantum theory, one can consult \cite{pt6}.
 A $\hat{\mathcal{P}}\hat{\mathcal{T}}$-symmetric Hamiltonian satisfies $[\hat{H},\hat{\mathcal{P}}\hat{\mathcal{T}}]=0$.  
Parity ($\hat{\mathcal{P}}$) operator changes the sign of position and momentum, i.e., $\hat{\mathcal{P}}: \hat{x}\to -\hat{x}$ and $\hat{\mathcal{P}}: \hat{p}\to -\hat{p}$. The antilinear time reversal operator acts as $\hat{\mathcal{T}}: \hat{x}\to \hat{x}$, $\hat{\mathcal{T}}: \hat{p}\to -\hat{p}$, and $\hat{\mathcal{P}}: \hat{i}\to -\hat{i}$.
  The consensus is that a quantum-mechanical interpretation of $\hat{\mathcal{P}}\hat{\mathcal{T}}$ symmetry may be implemented through a similarity transformation $\hat{H}_{\rho}=\hat{\rho}\hat{H}\hat{\rho}^{-1}$ of the $\hat{\mathcal{P}}\hat{\mathcal{T}}$ -symmetric Hamiltonian $\hat{H}$, such that $\hat{H}_\rho$ is self-adjoint ($\hat{H}_\rho^\dagger=\hat{H}_\rho$). In other words, $\hat{H}$ is quasi-Hermitian, i.e., Hermitian for a modified inner product \cite{pt52}. The existence of a  positive definite metric operator  to the construction of a consistent Rigged-Hilbert space corresponding to a $\hat{\mathcal{P}}\hat{\mathcal{T}}$ -symmetric Hamiltonian is fairly well accepted for time-independent metric operator\cite{pt61,pt62}. Since the present article deals with a Swanson oscillator with  time-dependent (TD) parameters, we reuire a TD metric operators. Mostafazadeh demonstrated that unitarity of the time evolution simultaneously with the observability of the Hamiltonian can not be guranteed for a TD metric operator \cite{TDmetric1}.  However, Fring and Moussa had shown that if the TD quasi-Hermiticity relation ($ \hat{H}^\dagger(t) \hat{\rho}(t) - \hat{\rho}(t) \hat{H}(t) = i\partial_t \hat{\rho}(t) $,\; with $ \hat{\rho}(t) = \hat{\eta}^\dagger \hat{\eta} $), and TD Dyson relation ($ \hat{H}_{\eta} = \hat{\eta}(t) \hat{H}(t) \hat{\eta}^{-1}(t) + i[\partial_t \hat{\eta}(t)] \hat{\eta}^{-1}(t) $) possess nontrivial solutions, then well defined observables and unitary time evolution can be achieved \cite{TDswansonmodel1,TDswansonmodel2}.
\\
Among the different domains of applicability of $\hat{\mathcal{P}}\hat{\mathcal{T}}$ -symmetric systems, optical systems are of great interest for their technological relevance \cite{pt7,pt8,pt9,pt10,pt11}. 
For instance, a multi-parameter family of the minimum-uncertainty squeezed states for the harmonic oscillator in nonrelativistic quantum mechanics and minimum-uncertainty squeezed states for atoms and photons
in a cavity are mathematically equivalent \cite{pts1}. Moreover, these types of systems are similar to the general form of the  Hamiltonian $\hat{H}_{pa}=\omega (\hat{a}^\dagger \hat{a} + \frac{1}{2}) +  \alpha \hat{a}^2 + \beta (\hat{a}^\dagger)^2$, which is a prototype class of  parametric amplification of simple harmonic oscillator with a quadratic field mode interaction. Time-independent system (constant parameters $\omega,\alpha,\beta$) of this type ($\hat{H}_{pa}$) is well studied for $\hat{\mathcal{P}}\hat{\mathcal{T}}$ -symmetric regime \cite{pth1,pth2,pth3}. Notably, both the Pt-symmetry phase and the non-PT symmetry phase have been studied in \cite{pth3}. Moreover, \cite{pth3} discussed the construction of inner products in both regions and in the boundary between them, as well as it deals with the study of Rigged Hilbert Space to construct bi-orthogonal sets in the non-PT-symmetry regime. Self-adjoint case  for time-dependent $\hat{H}_{pa}$, i.e., for $\alpha=\beta^*, \; \omega\in\mathbb{R}$ is explored in \cite{pts1}. However, there is a lack in the literature for the study of a general  $\hat{\mathcal{P}}\hat{\mathcal{T}}$ -symmetric time-dependent $\hat{H}_{pa}$.  A primary motivation of the present paper is to fulfill this gap. 
 \\
 We show that the entanglement behavior  of the time-dependent (TD) amplifiers can be destroyed with time-dependent parameters. The entanglement behavior has been
studied with the Wigner quasiprobability distributions (WQD), which provide a phase-space representation of quantum mechanics \cite{wqd1,wqd2,wqd3}. In particular, WQD is an attempt to interpret quantum mechanics as a statistical theory \cite{wqd1}. Being a phase-space formulation, WQD involves both real space and  momentum variables (distinctly different from SE).  Despite the philosophical debates, it is believed that one can conceptually identify where quantum corrections enter a problem by comparing it with the classical version \cite{wigner1,wigner2,wigner3}. The present article deals with the computational aspect of the time-variation of Wigner distribution for the prepared ground state of two spatially  separated $\hat{\mathcal{P}}\hat{\mathcal{T}}$-symmetric TD- oscillator moving in opposite directions to each other. 
 For time-independent case, the existence of a signature of entanglement at the origin of  phase-space for this type of cat-states is well known \cite{wigner4,wigner5,wigner6,wigner7,wigner8}. In this paper, we show that this signature is inevitable, even for a general time-dependent $\hat{\mathcal{P}}\hat{\mathcal{T}}$-symmetric oscillator cat-states. 
 \\
 To construct Wigner distribution $W(x,p,t) = \int_{-\infty}^{\infty} \psi^\dagger(x+x'/2) \psi(x-x'/2) e^{ipx'} dx' $, we need the solution of the Schr\"{o}dinger equation (SE) $\hat{H}\psi=i\partial_t\psi$. To solve TD-Schr\"{o}dinger equation corresponding to our TD-Hamiltonian, we have employed the Lewis-Riesenfeld (LR) method \cite{lrim0,lrim1,lrim2,lrim3,lrim4,lrim5}.  LR-method which states that for a system described by a TD Hamiltonian $\hat{H} (t)$, a particular solution of the associated Schr\"{o}dinger equation is given by the eigenstate $ \vert n,t \rangle $ of a TD-invariant operator $ \hat{\mathcal{I}} $ defined by $\partial_t\hat{\mathcal{I}} + (1/i) [\hat{\mathcal{I}}, \hat{H}] = 0$, apart from a TD-phase factor $ e^{i \theta_n(t)} $, where $ \theta_n(t) = \int_{0}^t \langle n,\tau\vert (i\partial_t -\hat{H}(t))\vert n,\tau\rangle  d\tau$ \cite{lrim6}. The construction of a LR-invariant operator of the form $\hat{\mathcal{I}}(t) = g_0(t) + g_1(t) \hat{p}^2 + g_2(t) \hat{x}^2 + g_3(t) \{ \hat{x}, \hat{p} \}$ is  equivalent to solving an integrable nonlinear Ermakov-Pinney (EP) equation \cite{Ermakovpenny1,Ermakovpenny2,Ermakovpenny3}, which makes the task nontrivial.
 Although the EP equation is integrable, only a few particular class of TD-parameters provides closed form solution of EP-equation \cite{Ermakovpenny4}. We illustrate this fact with a explicit closed form solution for a toy model. \\
 We transform the  quadratic invariant operator $\hat{\mathcal{I}}$ to a simpler (normal) form with the aid of a similarity transformation (symplectic group $Sp(2,\mathbb{R})$ \cite{sp1,sp2}). The symplectic diagonalization of $\hat{\mathcal{I}}$ enables us to factorize it in terms of annihilation and creation operators. The eigenstates of the annihilation operator provide the coherent states (CS). Construction of TD-phase factors (both the geometrical and dynamical \cite{phase1,phase2}) completes the task of obtaining the solution of the TD Schr\"{o}dinger equation. \\
 An explicit toy model is illustrated both in closed form solution analytically and graphically.
\section{PT symmetric oscillator and its Hermitian partner}
The Hamiltonian of a one-dimensional simple  harmonic oscillator reads
\begin{equation}\label{hamiltonianofsho}
	\hat{H}_{o}=\frac{1}{2m}\hat{p}^2 + \frac{1}{2}m\omega^2\hat{x}^2 = \hbar\omega(\hat{a}^\dagger \hat{a} + \frac{1}{2}),
\end{equation}
where the annihilation ($\hat{a}$) and creation operators ($\hat{a}^\dagger$) are given  in terms of the position ($\hat{x}$) and momentum ($\hat{p}$) operators as
\begin{equation}
	\hat{a}=\frac{1}{\sqrt{2}} ( l^{-1} \hat{x} + il\hbar^{-1}\hat{p}),\;\; \hat{a}^\dagger=\frac{1}{\sqrt{2}} ( l^{-1} \hat{x} - il\hbar^{-1}\hat{p}),\;\; [\hat{a},\hat{a}^\dagger]=1.  
\end{equation}
The TD length parameter $l(t)$ is defined by the TD effective mass $m(t)$ and frequency $\omega(t)$ of the oscillator as $l(t)=\sqrt{\frac{1}{m(t)\omega(t)}}$. We set $\hbar=1$, which will be followed throughout the present article. The commutation relation $
[\hat{a},\hat{a}^\dagger]=1
$ is independent of the TD parameters.
Let us introduce the following modulations of frequency and mass.
\begin{equation}\label{modulation}
	m\to m(1+\epsilon f(t)),\; \omega \to \omega(1+\frac{1}{2}\epsilon f(t)),
\end{equation} 
where the modulation amplitude $\epsilon$ is small enough so that we can neglect the terms  $\mathcal{O}(\epsilon^2)$ and higher.
Under modulation ~\eqref{modulation}, $\hat{H}_{o}$ changed to
\begin{equation}\label{modulatedhamiltonian}
\hat{H}_{o} \to \hat{H}_{mod}= \omega(\hat{a}^\dagger \hat{a} + \frac{1}{2}) + \frac{1}{2}\omega \epsilon f(t)(\hat{a}^2 +\hat{a}^{\dagger 2}).
\end{equation}
We wish to study a generalization of ~\eqref{modulatedhamiltonian} under the shade of non-Hermitian $\mathcal{P}\mathcal{T}$- symmetric Hamiltonian \cite{pt52}  
\begin{equation}\label{Hintermsofaadagger}
	\hat{H}=\omega (\hat{a}^\dagger \hat{a} + \frac{1}{2}) +  \alpha \hat{a}^2 + \beta (\hat{a}^\dagger)^2 .
\end{equation}
The Hamiltonian ~\eqref{Hintermsofaadagger} is previously studied for time-independent parameters \cite{pth1,pth2,pth3}. However, we are generalizing this into a TD parameter regime.
The interpretation of the TD -Hamiltonian ~\eqref{Hintermsofaadagger} is that it corresponds to a parametric amplification through anisotropic self-interaction of the field modes $\hat{a}$ and $\hat{a}^\dagger$. 
One can see that the Hamiltonian ~\eqref{Hintermsofaadagger} is non-Hermitian for the amplification parameters $\alpha(t)\neq\beta(t)$. However, if the real-valued parameters $\alpha(t)$, $\beta(t)$, and $\omega(t)$ are even functions of time, then  ~\eqref{Hintermsofaadagger} is $\mathcal{P}\mathcal{T}$-symmetric, which is evident from its following equivalent form.
\begin{equation}\label{Hintermsofxp}
\hat{H} = \frac{1}{2M} \hat{p}^2 + \frac{1}{2} M \Omega^2 \hat{x}^2 + \frac{i}{2}\nu_{-} \{ \hat{x}, \hat{p}\}.
\end{equation}
The notation $ \{ \hat{A}_1, \hat{A}_2 \} = \hat{A}_1 \hat{A}_2 + \hat{A}_2 \hat{A}_1 $ stands for the anticommutator of  $\hat{A}_1$ and $\hat{A}_2$.
\\
The equivalent mass $M(t)$, frequency $\Omega(t)$, and amplification parameter $\nu_{-}(t)$ are given by
\begin{equation}
M^{-1}(t) = (1-\nu_{+}/\omega)m^{-1},\; \Omega^2(t) = \omega^2 - \nu_{+}^2, \; \nu_{\pm}(t)=\alpha \pm \beta.
\end{equation}
For $\nu_{-}=0$, i.e., for $\alpha=\beta$, we have the hermitian counterpart of ~\eqref{Hintermsofaadagger} and ~\eqref{Hintermsofxp}. \\
We see that ~\eqref{Hintermsofxp} can be expressed in bilinear form as
\begin{eqnarray}
	\hat{H}= X^\dagger \mathcal{H}X ,\; \mbox{with} \; 
	\mathcal{H}=\left(\begin{array}{cc}
		h_{11} & h_{12}\\
		h_{21} & h_{22}
	\end{array}	\right),\; X=(X_1,X_2)^T (\hat{x},\hat{p})^T,\\
\mbox{with}\; h_{11}=\frac{1}{2}M\Omega^2,\; h_{22}= \frac{1}{2M},\; h_{12}=h_{21}=\frac{i}{2}\nu_{-}.
\end{eqnarray}
The notation $X^T$ stands for matrix transposition of $X$.  We note that the  intrinsic symplectic structure ($Sp(2,\mathbb{R})$) of the system is given by the commutation relations
\begin{equation}
	[X_j,X_l]=i J_{jl}; \; j,l=1,2.
\end{equation} 
$J_{ij}$ is the $ij^{\mbox{th}}$ element of the symplectic matrix $\hat{J}=i\hat{\sigma}_y$, where the notation $\sigma_i$ stands for Pauli matrices
\begin{eqnarray}
	\hat{\sigma}_x=\left(\begin{array}{cc}
		0&1\\
		1&0
	\end{array}\right), 
	\hat{\sigma}_y=\left(\begin{array}{cc}
		0&-i\\
		i&0
	\end{array}\right),\;
	\hat{\sigma}_z=\left(\begin{array}{cc}
		1&0\\
		0&-1
	\end{array}\right).
\end{eqnarray}
Since $\mathcal{H}$ is symmetric,   the symplectic eigenvalues of $\mathcal{H}$ are purely imaginary. In other words, the eigenvalues ($\lambda_{sp}$) of $\mathcal{H}_{sp}=J\mathcal{H}$ are purely imaginary ($\lambda_{sp}=\pm i\lambda_{h},\; \lambda_{h}\in\mathbb{R}$) \cite{Williamson1}.
Therefore the determinant of $\mathcal{H}_{sp}$ are always positive. In particular, the real spectrum of $\hat{H}$ corresponds to the constraint
\begin{equation}
	\omega^2-4\alpha\beta\ge 0,
\end{equation}
which fixes the parameter values for unbroken $\mathcal{P}\mathcal{T}$- symmetric domain, i.e., for which our $\mathcal{P}\mathcal{T}$- symmetric Hamiltonian has real eigenvalues. 
\\
Since $\hat{H}$ is not Hermitian, its eigenvectors are not orthogonal to each other in the usual inner product ($\langle u\vert v\rangle :=\int_\mathbb{R} u^*(x)v(x)dx $) of $\mathbb{L}^2$ space. Here $*$ denotes the complex conjugation.  To construct a modified inner product with an appropriate metric,  let us briefly summarize the proposal of \cite{TDswansonmodel1,TDswansonmodel2}.
Consider a mapping 
\begin{equation}
	\hat{\eta}(t): \vert\psi(t)\rangle \to \vert \phi(t)\rangle =\hat{\eta}(t)\vert \psi(t)\rangle,\; \mbox{with}\; \hat{\eta}^\dagger(t)=\hat{\eta}(t),
\end{equation}
which transforms the TD-Schr\"{o}dinger equation ($\hat{H}(t)\vert \psi(t)\rangle= i\partial_t\vert \psi(t)\rangle$) associated with the non-hermitian TD-Hamiltonian $\hat{H}$ to  the TD-Schr\"{o}dinger equation $\hat{H}_\eta(t)\vert \phi(t)\rangle = i\partial_t \vert \phi(t)\rangle$ associated with the  TD-Hamiltonian $\hat{H}_\eta$.  Hamiltonians are related via  TD-Dyson map
\begin{equation}\label{Hetareln}
	\hat{H}_{\eta} = \hat{\eta}(t) \hat{H}(t) \hat{\eta}^{-1}(t) + i[\partial_t \hat{\eta}(t)] \hat{\eta}^{-1}(t).
\end{equation}
Imposing the Hemiticity of $\hat{H}_\eta$, we get the TD quasi-Hermiticity relation
\begin{equation}\label{qusihermiticity}
\hat{H}^\dagger(t) \hat{\rho}(t) - \hat{\rho}(t) \hat{H}(t) = i\partial_t \hat{\rho}(t),\;\mbox{with}\;  \hat{\rho}(t) = \hat{\eta}^\dagger \hat{\eta}.
\end{equation}
~\eqref{qusihermiticity} implies that the TD probability densities in Hermitian and non-Hermitian systems are related through
\begin{equation}
	\langle \psi(t) \vert \tilde{\psi}(t)\rangle_\rho=\langle \phi(t)\vert \tilde{\phi}(t).
\end{equation}
For explicit construction, let us consider the ansatz
\begin{equation}\label{ansatzeta}
	\hat{\eta}=\exp(\hat{\Gamma}),\; \mbox{with}\; \hat{\Gamma}=\kappa_0 (\hat{a}^\dagger \hat{a} +1/2)+ \kappa \hat{a}^2 + \kappa^* (\hat{a}^\dagger)^2 .
\end{equation}
We would like to mention that the following constructions are mainly follow up of \cite{TDswansonmodel1,TDswansonmodel2}. One can see that
~\eqref{ansatzeta} can be decomposed into
\begin{equation}\label{etadecompose}
	\hat{\eta}=e^{\frac{1}{2}\mu_{+}a^{\dagger 2}} e^{\frac{1}{2}(\ln\mu_0)(\hat{a}^\dagger\hat{a} +1/2)} e^{\frac{1}{2}\mu_{+}^*a^{ 2}},
\end{equation}
where
\begin{eqnarray}
	\mu_{+}(t)= 2\kappa^* (\theta\coth\theta-\kappa_0)^{-1},\; \mu_0(t)= \frac{\theta^2 \vert \mu_{+}\vert^2  }{4\vert\kappa\vert^2}\mbox{cosech}^2\theta,\; \theta=\sqrt{\kappa_0^2 - 4 \vert \kappa\vert^2}.
\end{eqnarray}
Here, we consider $\kappa_0\in\mathbb{R}$ and $\kappa_0^2 - 4 \vert \kappa\vert^2 \ge 0$.
~\eqref{etadecompose} is convenient form for the time-derivative of $\hat{\eta}$. For convenience, we write 
\begin{equation}
	\mu_{+}= -\Phi_{\mu_{+}}e^{-i\phi_{\mu}},\; \mbox{with}\;\Phi_{\mu_{+}}^2 =\mu_{0}+\chi_{\mu},
\end{equation}
where
\begin{equation}
	\chi_\mu = \frac{\kappa_0+\theta\coth\theta}{\kappa_0 -\theta\coth\theta},\;\; \kappa/\kappa_0=\vert\kappa/\kappa_0\vert e^{i\phi_\mu}.
\end{equation}
Using ~\eqref{ansatzeta},~\eqref{etadecompose} and ~\eqref{Hintermsofaadagger} in ~\eqref{Hetareln},  one obtains 
\begin{equation}\label{Hrho}
	\hat{H}_{\eta} =   \omega_0(\hat{a}^\dagger \hat{a}+1/2)+ \alpha_0 \hat{a}^2 +\beta_0  \hat{a}^{\dagger2},
\end{equation}
with the TD coefficients 
\begin{eqnarray}
\omega_0 &=& - \left[\omega(\chi_\mu+\vert\mu_{+}\vert^2 ) + 2 ( \alpha \mu_{+} + \beta \mu_{+}^*\chi_\mu) -i (\dot{\mu}_0- 2\mu_{+}\dot{\mu}_{+}^*)/2\right]/\mu_{0},\\
\alpha_0 &=& (\alpha + \omega \mu_{+}^* + \beta \mu_{+}^{*2} + i\dot{\mu}_{+}^*/2)/\mu_{0}, \\
\beta_{0} &=& \left[\omega\chi_\mu \mu_{+} + \alpha\mu_{+}^2 +\beta\chi_{\mu}^2  + i(\mu_{0}\dot{\mu}_{+} + \mu_{+}^2 \dot{\mu}_{+}^*- \mu_{+}\dot{\mu}_{0})/2\right]/\mu_{0} .
\end{eqnarray} 
Here dot denotes the derivative with respect to time. 
Since we want $\hat{H}_\eta$ to be Hermitian, we need $\omega_0\in \mathbb{R}$ and $\alpha_0=\beta_0^*$, which fix the free parameters $\kappa$ and $\kappa_0$ accordingly. In particular,
\begin{eqnarray}
\dot{\mu}_0 = 2\vert \omega \vert (\chi_{\mu}+\Phi_{\mu_{+}}^2)\sin\phi_{\omega} + 2\Phi_{\mu_{+}} (\dot{\Phi}_{\mu_{+}} + 2\vert \alpha\vert \sin(\phi_{\mu}-\phi_{\alpha}) - 2\vert \beta\vert \chi_{\mu}\sin(\phi_{\mu}+\phi_{\beta})), \label{diff1}\\
(\chi_{\mu}-1)\dot{\Phi}_{\mu_{+}}/2 = (\vert\omega\vert \Phi_{\mu_{+}}\sin\phi_{\omega}+ \vert \alpha\vert \sin(\phi_\mu-\phi_{\alpha}))(1-\Phi_{\mu_{+}}^2) \nonumber \\ + \vert \beta \vert ((2\chi_{\mu}-1)\Phi_{\mu_{+}}^2-\chi_{\mu}^2)\sin(\phi_{\mu}+\phi_{\beta}), \label{diff2}\\
(\chi_{\mu}-1)\Phi_{\mu_{+}}\dot{\phi}_\mu/2= \vert \alpha \vert (1-\Phi_{\mu_{+}}^2)\cos(\phi_{\mu}-\phi_{\alpha}) + \vert \beta\vert (\Phi_{\mu_{+}}^2-\chi_{\mu}^2) \cos(\phi_{\mu}  + \phi_{\beta}) ) \nonumber \\ + \vert\omega\vert (\chi_{\mu}-1)\Phi_{\mu_{+}}\cos\phi_{\omega}.\label{diff3}
\end{eqnarray}
Here $\phi_\alpha,\phi_\beta$ and $\phi_\omega$ are phases of $\alpha,\beta$ and $\omega$, respectively.
It is not obvious that the set of coupled nonlinear differential equations ~\eqref{diff1}-~\eqref{diff3} admits any solution. However, for special restriction on parameters it might. For instance, let us consider the following special choices.
\subsection{Special case ($\alpha,\beta,\omega$ real): } For real valued $\alpha,\beta,\omega$, the coupled nonlinear equations simplified to followings.
\begin{eqnarray}
	\frac{d}{dt}(\mu_0-\Phi_{\mu_{+}}^2) &=& 4\Phi_{\mu_{+}}(\alpha-\beta\chi_{\mu})\sin\phi_{\mu}, \\
(\chi_{\mu}-1)\dot{\Phi}_{\mu_{+}}/2 &=& (\alpha(1-\Phi_{\mu_{+}}^2)+ \beta((2\chi_{\mu}-1) \Phi_{\mu_{+}}^2 -\chi_{\mu}^2))\sin\phi_{\mu}, \\
(\chi_{\mu}-1)\Phi_{\mu_{+}}\dot{\phi}_\mu &=& 2(\alpha(1-\Phi_{\mu_{+}}^2)+\beta (\Phi_{\mu_{+}}^2-\chi_{\mu}^2))\cos\phi_\mu + 2\omega (\chi_{\mu}-1)\Phi_{\mu_{+}}.
\end{eqnarray}  
\subsection{Special case ($\kappa$ real):} We had considered real valued $\kappa_0$. If we further restrictcs ourselves on the domain of real valued $\kappa$, i.e., $\phi_\mu=0$, the constraint equations reduced to
\begin{eqnarray}
	\Phi_{\mu_{+}}=\phi_{\mu_{0+}}=\mbox{constant},\; \mu_0=\mu_{00}=\mbox{constant},\;
	\chi_{\mu}= 1-\omega/\beta.
\end{eqnarray} 
Here we have discarded the another possibility of $\chi_{\mu}=-1$, which implies $\kappa_0=0$. To get a nontrivial constraint, we are only considering here $	\chi_{\mu}= 1-\omega/\beta$, which immediately implies
\begin{equation}\label{constraintparameterequation}
	\omega\kappa_0\tanh\theta=\theta(\omega-2\beta).
\end{equation}
\subsection{Toy Model:}
To appreciate that the constraint equation ~\eqref{constraintparameterequation} indeed admits a real solution, let us consider the toy model 
\begin{equation}
	\omega=2\beta_{00},\; \beta=\beta_{00}\cos(\phi_{00}+\omega_{00}t),
\end{equation}
where $\beta_{00}, \omega_{00},\phi_{00}$ are constants. This leads to
\begin{equation}\label{constraint2}
	\kappa_0 \tanh\theta=2\theta \sin^2((\phi_{00}+\omega_{00}t)/2).
\end{equation}
If we consider the extreme special case, namely 
\begin{equation}
	\theta\to 0,\; \mbox{i.e.}, \kappa_0\to 2\kappa,
\end{equation}
then ~\eqref{constraint2} is reduced to
\begin{equation}
	\kappa_0= 2 \sin^2((\phi_{00}+\omega_{00}t)/2).
\end{equation}
Thus we see that for specific choice of parameter values we can have a viable hermitian hamiltonian corresponding to a time-dependent $PT$-symmetric system.
\\
From now on we shall consider $\alpha_0(t)$ is real valued. 
Essentially what follows is that the Hermitian Hamiltonian in $\eta$-space corresponds to a TD-ordinary harmonic oscillator with the Hamiltonian
\begin{equation}
	\hat{H}_\rho (t)= \frac{1}{2M_0}\hat{p}^2 + \frac{1}{2}M_0 \Omega_0^2 \hat{x}^2 ,
\end{equation}
with the following effective TD-mass $	M_0(t)$ and TD-frequency $\Omega_0(t)$.
\begin{equation}
	M_0=m(1-\epsilon_\omega)^{-1},\; \Omega_0=\omega\sqrt{1-\epsilon_\omega^2},\; \mbox{where}\; \epsilon_\omega(t)= 2\alpha_0/\omega.
\end{equation}
\section{Lewis-Riesenfeld invariant operator and the solution of Schr\"{o}dinger equation}
The set of operators $ \mathcal{S}_{\mathcal{O}} = \{ \hat{x}^2, \hat{p}^2, \{ \hat{x}, \hat{p} \} \} $ form a closed quasi-algebra with respect to $\hat{H}_\rho$. Accordingly, we choose an ansatz for a  time-dependent invariant operator of the form
\begin{equation}
	\hat{\mathcal{I}}(t) = g_0 + g_1(t) \hat{p}^2 + g_2(t) \hat{x}^2 + g_3(t) \{ \hat{x}, \hat{p} \}.
\end{equation}
The  invariance condition
\begin{equation}\label{LRIcondition}
	\partial_t \hat{\mathcal{I}}-i[\hat{\mathcal{I}},\hat{H}_\rho] =0,
\end{equation}
 is equivalent to the following set of coupled equations.
\begin{eqnarray}\label{LRIMparametereqn}
\dot{g}_0=0,\;	\dot{g}_1 = -(2/M_0)g_3,\; \dot{g}_2=2M_0\Omega_0^2g_3,\; \dot{g}_3=M_0\Omega_0^2 g_1 - (1/M_0)g_2.
\end{eqnarray}
Here dot denotes the total derivative  with respect to time. We shall set the arbitrary additive constant $g_0=0$. The set of equations ~\eqref{LRIMparametereqn} corresponds to the invariance 
\begin{equation}\label{LRIMparaminvariance}
g_3^2(t) - g_2(t)g_1(t)= -\eta_0^2 =\mbox{constant}.
\end{equation}
We define an auxiliary variable $g_1(t)=\eta^2(t)$, which enables us to rewrite equations ~\eqref{LRIMparametereqn} and ~\eqref{LRIMparaminvariance} as
\begin{eqnarray}
	g_1=\eta^2,\; g_3=-M_0\eta\dot{\eta},\; g_2=M_0^2\dot{\eta}^2 + \eta_0^2/\eta^2 ,
\end{eqnarray}
along with the following Ermakov-Pinney (EP) equation \cite{Ermakovpenny1,Ermakovpenny2,Ermakovpenny3}.
\begin{equation}\label{ermakovpinney}
	\ddot{\eta} + (\dot{M}_0/M_0) \dot{\eta} + \Omega_0^2 \eta = \eta_0^2/(M_0^2\eta^3).
\end{equation}

To solve the eigenvalue equation of $\hat{\mathcal{I}}(t)$, namely
\begin{equation}
\hat{\mathcal{I}} \Phi = \lambda\Phi,
\end{equation}
we first observe that $\hat{\mathcal{I}}$ can be expressed in the bilinear form
\begin{eqnarray}
	\hat{\mathcal{I}}=X^\dagger \hat{\mathcal{H}}_{\mathcal{I}}X,\; \mbox{with}\; 
		\hat{\mathcal{H}}_{
			\mathcal{I}}=\left(\begin{array}{cc}
			g_2 & g_3\\
			g_3 & g_1
		\end{array}\right)
\end{eqnarray}
Moreover, $\hat{\mathcal{I}}$ holds the  symplectic structure ($Sp(2,\mathbb{R})$)
\begin{eqnarray}
 \left[i\hat{\mathcal{I}},X\right]=\hat{\Lambda}X,\;
\; \mbox{with} \;\;\hat{\Lambda}=J \hat{\mathcal{H}}_\mathcal{I}.
\end{eqnarray}
To diagonalize $\hat{\Lambda}$, we first note that it has two purely imaginary eigenvalues  $\lambda_\pm = \pm i\eta_0 $. If the eigenvalue $-i\eta_0$ corresponds to a left eigenvector $u_{-}$ of $\hat{\Lambda}$, then the left eigen-vector corresponding to $i\eta_0$ is given by $u_+=u_{-}^*$. Accordingly, the right vectors $v_{\mp}$  corresponding to the eigenvalues $\mp i\eta_0$ are given by $v_{-}=-\sigma_y u_{-}^\dagger$, and $v_{+}=v_{-}^*$. The normalization can be fixed from the orthonormalization conditions $u_{-}v_{-}=u_{+}v_{+}=1,\; u_{-}^*v_{+}=u_{+}v_{-}^*=0$. By defining the similarity transformation matrix $Q=(v_{-},v_{+})$ and $Q^{-1}=(u_{-}^T,u_{-}^{*T})^T$, we obtain a diagonal representation of $\hat{\Lambda}$. In particular,
\begin{eqnarray}
	u_{-}=(u_{11},u_{12})=\frac{1}{\sqrt{2\eta_0 g_1}} (g_3-i\eta_0,g_1), \; \Lambda_D=Q^{-1}\Lambda Q = \mbox{diag}(-i\eta_0,i\eta_0).
\end{eqnarray}
If we define a normal coordinate by
\begin{equation}
	\hat{a}_{-}=u_{-}X,\;\hat{a}_{+}=u_{-}^*X,\; \mbox{i.e.}, A= (\hat{a}_{-},\hat{a}_{+})^T = Q^{-1}X,
\end{equation}
one can see that $\hat{a}_{-}$ and  $\hat{a}_{+}$ satisfy the following criterion to be a bonafide annihilation and creation operators, respectively.
\begin{equation}
\hat{a}_{+}=\hat{a}_{-}^\dagger,\; \; [\hat{a}_{-},\hat{a}_{+}]=1.
\end{equation}
The inverse relations are also immediately given by
\begin{equation}\label{xpinaadagger}
	\hat{x}=iu_{12}(\hat{a}_{-}-\hat{a}_{+}),\; \hat{p}= i(u_{11}\hat{a}_{+}-u_{11}^* \hat{a}_{-}).
\end{equation}
What follows is that the invariant operator $\hat{\mathcal{I}}$ can be factorized in terms of the annihilation can creation operators as
\begin{eqnarray}
	\hat{\mathcal{I}}&=&A^\dagger Q^\dagger \hat{H}_{\mathcal{I}}QA= iA^\dagger \sigma_z \Lambda_D A \;\;\; (\because Q^\dagger=-\sigma_zQ^{-1}\sigma_y) \\	
	&=& A^\dagger \mbox{diag}(\eta_0,\eta_0)A= \eta_0 A^\dagger A = 2 \eta_0 (\hat{a}_{-}^\dagger \hat{a}_{-}+1/2).
\end{eqnarray}
That means the eigenstates of TD-operator $\hat{\mathcal{I}}$ can be written in terms of the orthonormal eigenstates of the number operator as follows.
\begin{equation}
\hat{\mathcal{I}}\vert n\rangle =\epsilon_n\vert n\rangle,\;\; \epsilon_n = 2\eta_0 (n+1/2),\; n=0,1,2,....,
\end{equation}
where the $n^{th}$ state $\vert n \rangle$ is given by the ground state $\vert 0\rangle$ through
\begin{equation}
\vert n\rangle = \frac{1}{\sqrt{n!}} (\hat{a}_{-}^\dagger)^n \vert 0\rangle,\;\;  \hat{a}_{-} \vert 0\rangle =0,\;\; \langle n \vert m\rangle =\delta_{nm} .
\end{equation}
The symbol $\delta_{ij}$ stands for Kronecker delta, whose value is one for equal indices and zero otherwise.
In position representation ($\{\vert x\rangle\}$), the ground state ($\phi_0(x,t)$) eigenfunction of $\hat{\mathcal{I}}$ can be written as
\begin{equation}
	\phi_0(x,t) =\frac{\sqrt[4]{\eta_0/\pi}}{\sqrt{\eta}} \exp(-\eta^{-2}(\eta_0 -iM_0\eta\dot{\eta})x^2/2),
\end{equation}
provided $\eta_0>0$ in order to ensure the normalization. From now on, without loss of generality, we shall consider $\eta_0=1$.
\\
According to LR-invariant theorem, the solution of the TD-Schr\"{o}dinger equation 
$
\hat{H}_\rho \psi =i\partial_t \psi,
$
is given by the eigenstates of LR-invariant operator through
\begin{equation}
	\psi (t)= \sum_{n=0}^{\infty}c_n \vert n\rangle e^{i\theta_{n}(t)}, 
\end{equation}
where the TD-phase factors $\theta_n(t)$ have to be determined from the consistency condition 
\begin{equation}
	\theta_n(t)= \int_{0}^{t} \langle n,\tau \vert [i\partial_\tau -\hat{H}_\rho ]\vert n,\tau \rangle d\tau,\;\; \mbox{and}\; c_n=\langle n\vert \psi(0)\rangle.
\end{equation} 
One can split $\theta_n(t)$ into the sum $\theta_n=\theta_n^{(g)}+ \theta_n^{(d)}$ of a geometric phase $\theta_n^{(g)}$ (including the Berry phase phenomenon for slowly varying parameters), and a dynamical phase $\theta_n^{(d)}$, respectively, as follows.
\begin{equation}
\dot{\theta}_n^{(g)}=i\langle n,t \vert \partial_t \vert n,t\rangle ,\;
\dot{\theta}_n^{(d)}=-\langle n,t \vert \hat{H}_\rho \vert n,t \rangle .
\end{equation}
To compute $\theta_n^{(d)}$, we first express $\hat{H}_\rho$ in terms of $a_{\pm}$ as
\begin{equation}
	\hat{H}_\rho = \omega_\rho (\hat{a}_{-}^\dagger \hat{a}_{-}+1/2) - \alpha_\rho \hat{a}_{-}^2 - \alpha_\rho^* \hat{a}_{+}^{ 2},
\end{equation}
where
\begin{eqnarray}
	\omega_\rho (t)=\frac{1}{2M_0\eta^2} (1  +  M_0^2 \Omega_0^2 \eta^4 + M_0^2 \eta^2 \dot{\eta}^2),\; 
	\alpha_\rho (t)= \frac{1}{2}(\omega_\rho - \frac{1}{M_0\eta^2}-\frac{i\dot{\eta}}{\eta}).
\end{eqnarray}
Accordingly the dynamical phase reads
\begin{equation}
	\dot{\theta}_n^{(d)}= -(n+1/2)\omega_\rho \in \mathbb{R}.
\end{equation}
Since $\hat{a}_{-}$ does not commute with $\partial_t\hat{a}_{-}$ for $\eta_0\neq 0$ ( $\eta_0=0$ is trivial situation), the straightforward computation of $\langle n,t\vert \partial_t \vert n,t \rangle$ is not immediate. Rather, we can use a representation to express the states in their explicit form and compute the required quantities. In position $(\vert x\rangle)$ representation, we have
\begin{equation}
	\phi_n(x,t)= \langle x\vert n,t\rangle = \frac{ \sqrt[4]{g/\pi}}{\sqrt{2^n n!}} e^{-gx^2/2} H_n(\sqrt{g}\;x),\; \mbox{with}\; g(t)=(1 +ig_3)/g_1.
\end{equation}
Using the property $H_n'(x)=2nH_{n-1}(x)$ and the orthogonality of Hermite polynomials ($H_n(x)$), one can evaluate
\begin{equation}
	\dot{\theta}_n^{(g)}=n\dot{g}\sqrt{g_1/(2 g)} + i \dot{g}/(4g)-i (n+1/2)\dot{g} g_1/2.
\end{equation}
Here prime denotes the derivative with respect to $x$.\\
Calculation of physically relevant properties (e.g., expectation values) requires the inner product, hence the adjoint of the state. On the other hand, $\theta_n$'s appear in the phase factor. Hence, it is enough to concentrate on the imaginary part of $\theta_n^{(g)}$ for  the computation of relevant quantities.  In particular
\begin{eqnarray}
\dot{\theta}_{n,im}^{(g)}=	\Im(\dot{\theta}_n^{(g)}) = \frac{n}{2\sqrt{g_1}} [\dot{g}_3r_{+}/r_g - \dot{g}_1 g_2 r_{-}] \eta+ \frac{1}{8}\frac{d}{dt}\ln(g_2/g_1) + \frac{1}{2}(n+1/2) \dot{g}_1/g_1.
\end{eqnarray}
where $r_g=\sqrt{g_1g_2},\; r_{\pm}=\sqrt{1\pm 1/r_g}$. It is easy to note that $r_g^2 > 1$. Now the total solution of the TD-Schr\"{o}dinger equation reads
\begin{eqnarray}\label{tdsesoln}
	\psi(x,t)= \sum_{n=0}^{\infty} c_n \psi_n(x,t) = \sum_{n=0}^{\infty}  \frac{c_n  e^{-n \int_{0}^{t} (\dot{g}_3 r_{+} /r_g - \dot{g}_1 g_2 r_{-})d\tau }}{\eta^{2n} \sqrt[4]{1 + M_0^2 \eta^2 \dot{\eta}^2}}	\phi_n(x,t) . 
\end{eqnarray}

However, not all the parameter values are allowed.  Every bonafide variance matrix ($\hat{\mathcal{V}}$) has to satisfy the following natural constraint imposed through the Robertson-Schr\"{o}dinger uncertainty principle (RSUP) \cite{rsup2}.
\begin{equation}
\hat{\mathcal{V}}+ i \hat{J}/2 \ge 0,
\end{equation}
where the variance elements ($\mathcal{V}_{ij}$) are  defined by 
\begin{equation}
	\mathcal{V}_{ij}= \frac{1}{2} \langle \{ \hat{X}_i, \hat{X}_j \}\rangle -\langle  \hat{X}_i\rangle \langle  \hat{X}_j\rangle .
\end{equation}
For the system under study, the explicit forms of the elements of $\hat{\mathcal{V}}$ read as
\begin{eqnarray}
	\mathcal{V}_{11}= \frac{g_1}{2}(2n+1) e^{-2\theta^{(g)}_{n,im}},\;
		\mathcal{V}_{22}= \frac{g_2}{2}(2n+1) e^{-2\theta^{(g)}_{n,im}},\;
		\mathcal{V}_{12}=\frac{M_0 \dot{\eta}}{2}(2n+1) e^{-2\theta^{(g)}_{n,im}}.
\end{eqnarray}
For the ground state ($n=0$), one can evaluate the exact form 
\begin{equation}
\theta^{(g)}_{0,im} = \frac{1}{8} \ln (g_1g_2),
\end{equation} 
 which enables us to express RSUP with the following constraint equation.
 \begin{equation}
 	1+M_0^2\eta^2\dot{\eta}^2 \ge M_0^2\dot{\eta}^2 + \sqrt{	1+M_0^2\eta^2\dot{\eta}^2}.
 \end{equation}

\section{Wigner phase-space distribution}
The characteristic function $\hat{\mathcal{M}}(\tilde{\tau},\tilde{\theta})$ of a random variable  $X=(\hat{x},\hat{p})$ is defined by $\hat{\mathcal{M}}(\tilde{\tau},\tilde{\theta})= e^{i\mbox{Trace}(\tilde{t}^T X)},$ with the parameter $\tilde{t}=(\tilde{\tau},\tilde{\theta})$.
The Fourier transformation of the expectation value of the characteristic function ($\hat{\mathcal{M}}$) is known as the Wigner quasiprobability distribution (WQD) \cite{wqd1,wqd2,wqd3}. 
\\
Let us consider that we prepare the system at the ground state $\psi_0(x,t)$ (~\eqref{tdsesoln}) of the oscillator at $t=0$. Momentum space wavefunction corresponding to the prepared ground state of the TD-oscillator (in position representation) centered at $x=x_0$ is
\begin{eqnarray}\label{psitildep}
	\tilde{\psi}_0(p,t)&=&\frac{1}{\sqrt{2\pi}}\int_{-\infty}^{\infty}\psi_0(x-x_0,t)e^{-ipx}dx \\
	 &=& (k_g/\sqrt{g}) e^{-\frac{1}{2g}p^2 -ipx_0},\;\mbox{with}\; k_g=\sqrt[4]{\frac{g}{\pi \eta^4 \vert g\vert^2}} .
\end{eqnarray}
In order to study the entanglement behavour let us consider the following composite wave function which is composed of two unpolarized ground state (prepared) of the form ~\eqref{psitildep}.
\begin{equation}\label{psicomposite}
	\tilde{\psi}_c (p,t)= \frac{1}{\sqrt{2}}[\tilde{\psi}_{0}(p-p_0,x_0)+\tilde{\psi}_{0}(p+p_0,-x_0)].
\end{equation}
This is thus  one portion of the wave function centered at $p_0$
and moving in the positive $p$ direction, and a second portion
centered at $-p_0$ and moving in the negative $p$ direction.  The density operator in momentum space is defined by
\begin{equation}\label{densityoperator}
	\tilde{\rho}_w (p,p')= \tilde{\psi}_c (p-p'/2) \tilde{\psi}_c^\dagger (p+p'/2).
	\end{equation}
Fourier transformation (FT) of the density operator over $p'$ provides the Wigner phase-space distribution $(W(x,p,t))$ of the system. In particular,
\begin{equation}\label{wignerdefn}
	W(x,p,t)=\frac{1}{\sqrt{2\pi}} \int_{-\infty}^{\infty}  	\tilde{\rho}_w (p,p') e^{ip'x}dp'.
\end{equation}
Using ~\eqref{psitildep} in ~\eqref{psicomposite} one can construct ~\eqref{densityoperator} in straightforward manner. After performing the FT as in ~\eqref{wignerdefn}, we get the following explicit form for the  Wigner phase-space distribution function.
\begin{equation}
	W(x,p,t)=\sqrt{\frac{g_r}{2\pi \vert g\vert}}(I_1+I_2+I_3+I_4).
\end{equation}
Where
\begin{eqnarray}
	I_1=  \exp[ -(p-p_0)^2/g_r -\vert g\vert^2 (x+x_0)^2/g_r + 2g_i (p-p_0)(x+x_0)/g_r].\\
	I_4= \exp[ -(p+p_0)^2/g_r -\vert g\vert^2 (x-x_0)^2/g_r + 2g_i (p+p_0)(x-x_0)/g_r].\\
	I_2+I_3= 2 \exp[-p^2/g_r -g_rx^2  - g_i (g_i x^2 -2xp)/g_r] \cos(2x_0p+2px_0).
\end{eqnarray}
\section{Toy Model}
The nonlinear nature of the Ermakov Pinney equation suggests that it is difficult (if not impossible) to obtain solutions of ~\eqref{ermakovpinney} for even a simple choice to TD-mass and frequency. Only a judicious choice will lead to a closed-form solution.
For instance for the following parameter choice
\begin{equation}\label{choiceM0Omega0}
	M_0(t)=\eta_0 t,\; \Omega_0(t)=\eta_0/t,
\end{equation}
~\eqref{ermakovpinney} admits closed form solutions. In particular for $\eta_0=1$, the choice ~\eqref{choiceM0Omega0} enables to write the closed form solutions of ~\eqref{ermakovpinney} as
\begin{eqnarray}
	\eta_{j\pm}(t)= (-1)^j \sqrt{c_1 \pm \sqrt{(c_1^2-1)\zeta/(1+\zeta)}},\; j=1,2.\\
	\mbox{with}\;\;	\zeta= \tan^2(2c_2-2\ln t).
\end{eqnarray}
The allowed value of the integration constants $c_1$ and $c_2$ have to be fixed from consistency. 
\begin{center}
	\begin{figure}[!h]
		\centering\includegraphics[totalheight=6cm]{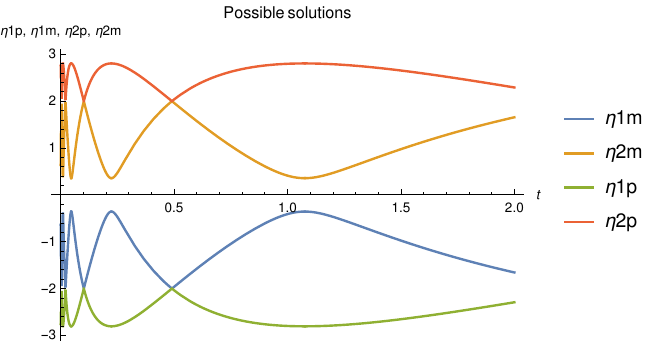}\\
		\caption{Symmetry of the equation ~\eqref{ermakovpinney} implies if  $\eta$ is a solution, then $-\eta$ is also a solution (rare property, indeed for a nonlinear equation). Two independent solutions $\eta 1p=\eta_{1+}$ and $\eta2p=\eta_{2+}$ (or equivalently $\eta 1m=\eta_{1-}=-\eta 1p$ and $\eta2m=\eta_{2-}=-\eta 2p$) are shown in the figure. The integration constants are set to the value $c_1=c_2=4$ for scalling purpose.}\label{sampleFig3}
	\end{figure}
\end{center}
 FIG.~\ref{sampleFig3} represents the solution for $\eta_0=1$. For our practical purpose of the problem under consideration, we shall choose only positive solutions, namely $\eta1p$ and $\eta2p$.
\begin{center}
	\begin{figure}[!h]
		\centering\includegraphics[totalheight=6cm]{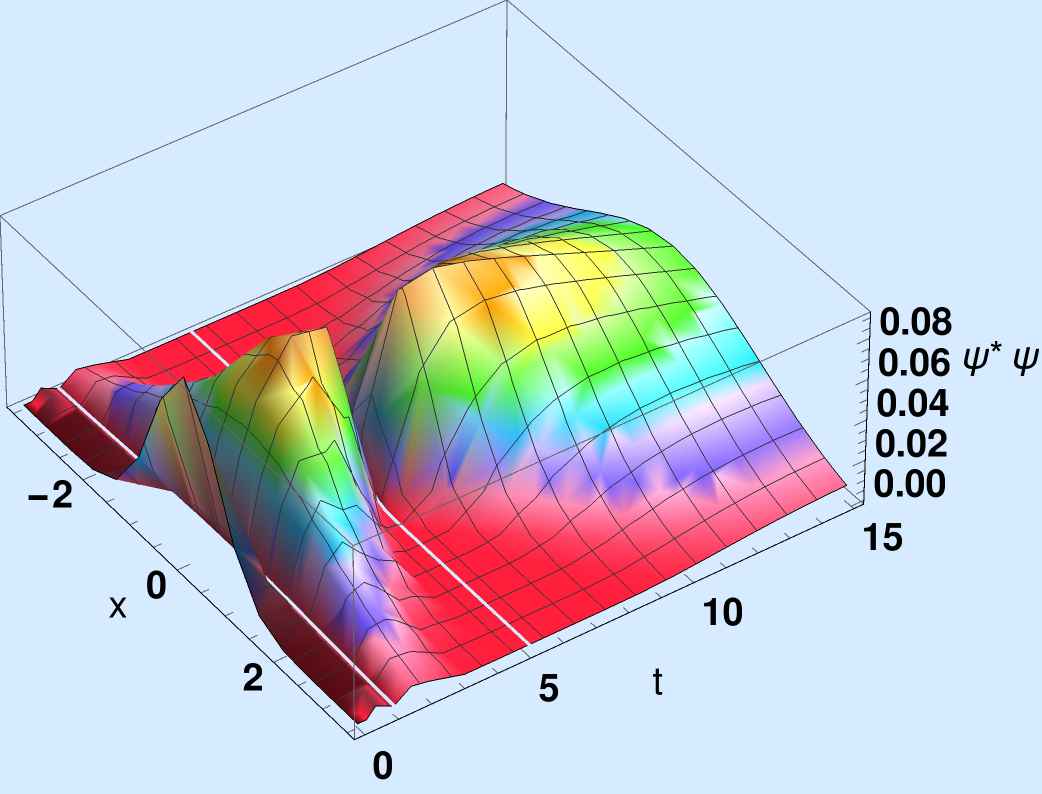}\\
		\caption{Probability density corresponding to $\psi_0(x,t)$. }\label{sampleFig4}
	\end{figure}
\end{center}
FIG.~\ref{sampleFig4} represents the probability density $\rho(x,t)=\psi^*(x,t)\psi(x,t)$ corresponding to the ground state wave function $\psi_0(x,t)$ for the choice of parameters ~\eqref{choiceM0Omega0}.

\begin{center}
\begin{figure}[!h]
	\centering\includegraphics[totalheight=6cm]{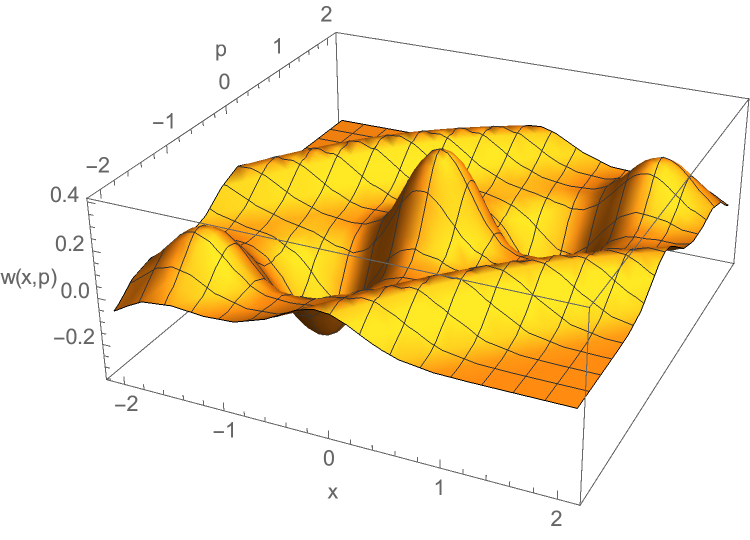}\\
	\caption{Wigner distribution W(x,p) for time $t\to 0.1$. }\label{sampleFig5}
\end{figure}
\begin{figure}[!h]
	\centering\includegraphics[totalheight=6cm]{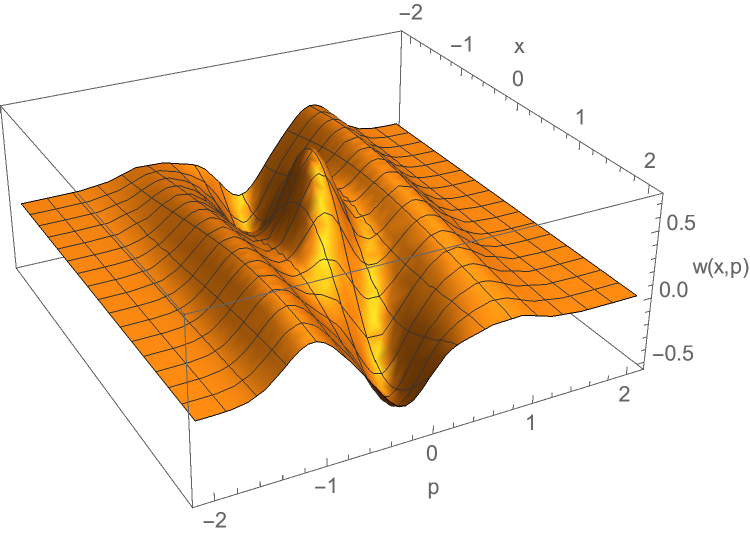}\\
	\caption{Wigner distribution W(x,p) for time $t\to1$. }\label{sampleFig6}
\end{figure}
\begin{figure}[!h]
	\centering\includegraphics[totalheight=6cm]{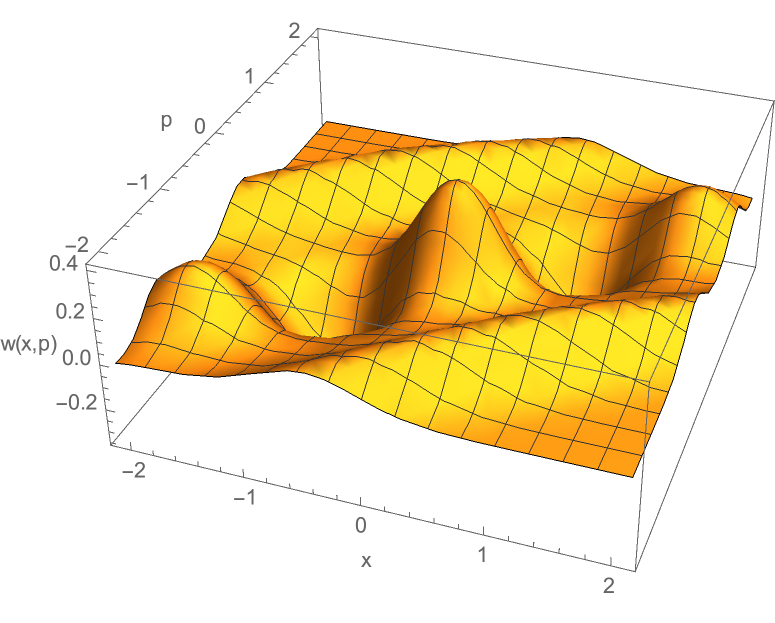}\\
	\caption{Wigner distribution W(x,p) for time $t\to2$. }\label{sampleFig7}
\end{figure}
\begin{figure}[!h]
	\centering\includegraphics[totalheight=6cm]{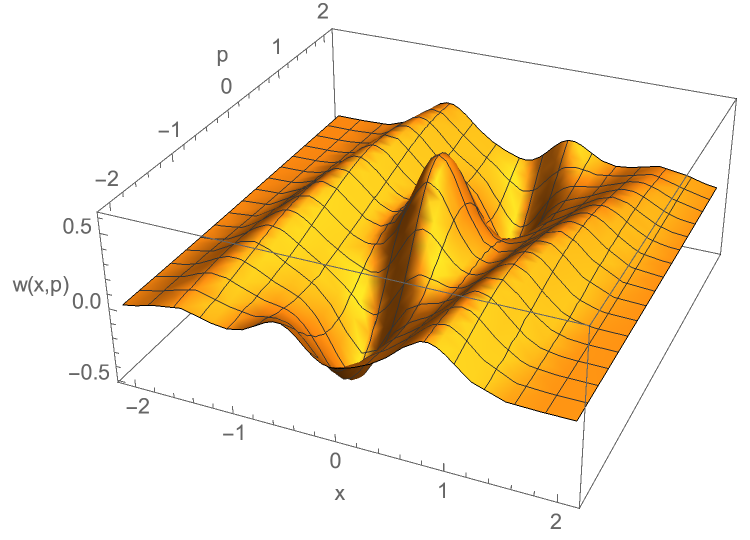}\\
	\caption{Wigner distribution W(x,p) for time $t\to 100$ }\label{sampleFig8}
\end{figure}
\begin{figure}[!h]
	\centering\includegraphics[totalheight=6cm]{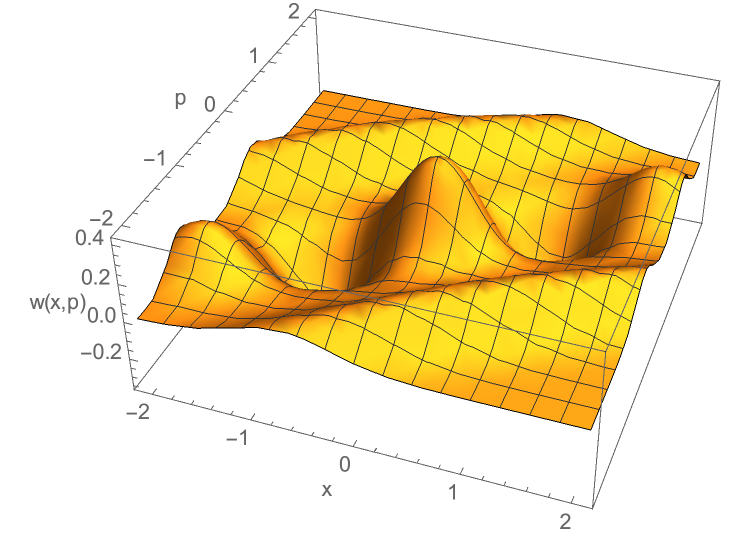}\\
	\caption{Wigner distribution W(x,p) for time $t\to 1000$ }\label{sampleFig9}
\end{figure}
\begin{figure}[!h]
	\centering\includegraphics[totalheight=6cm]{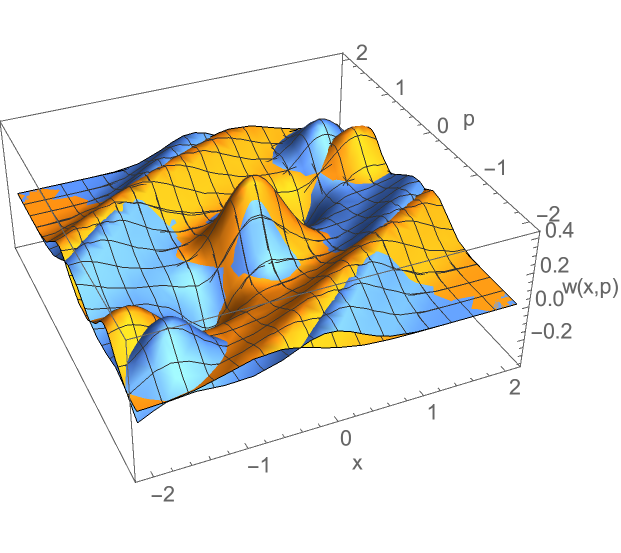}\\
	\caption{Wigner distribution W(x,p) for time $t\to 0.1$ and $t\to 1000$ for comparison.  }\label{sampleFig10}
\end{figure}
\end{center}
Wigner distribution ($W(x,p)$) corresponding to the same parameter choice (~\eqref{choiceM0Omega0}) have been illustrated on  FIG.~\ref{sampleFig5} to FIG.~\ref{sampleFig10}. FIG.~\ref{sampleFig5}, FIG.~\ref{sampleFig6} and FIG.~\ref{sampleFig7} corresponds to the distribution function for  time $t\to 0.1,t\to1$, and $ t\to 2$, respectively. After large time ($t\to 100, t\to 1000$) distribution looks like FIG.~\ref{sampleFig8} and FIG.~\ref{sampleFig9}. FIG.~\ref{sampleFig10} represents the relative variation of $W(x,p)$ for time $t\to0.1$ and $t\to 1000$. One can observe that the antanglement behavour always present at origin. In other words, tarce of their history is always remains even for time-dependent (TD) parameters. Once they are entangled, they are entangled forever, which can not be destroyed by  TD-parameters (at least for our parameter choice).
\section{Conclusions}
The measure of the number of separable states over  the entangled states is negligible \cite{entanglemeasure1,entanglemeasure2}. In particular, the presence of $\ln\ln d$ in the upper bound of the separable states in $d=2^N$-dimensional Hilbert space  of $N$-spin $1/2$-particles (qubit) \cite{entanglemeasure1}, makes one curious whether the number of separable states has a connection with the number of prime numbers in the set of integers (recall that the infinite series consisting of the inverse of prime numbers diverges as $\ln\ln d$). The present study of the time-dependent (TD) wigner distribution  supports the fact that in fact,  it is hard to find out separable states. In particular, if two states are entangled at some time, then they are entangled forever. Even a TD parameter (at least for the parameter choice under consideration in the toy model) can not destroy it.\\
We have explicitly constructed the closed form of Wigner distribution for TD- parametric amplifiers. This form can be used to compute the joint characteristic function for the position and momentum of the oscillator through inverse Fourier transformation. Thus the theoretical probability distribution for  various phases of the local oscillator for a homodyne tomography can be tallied with the measured distribution for different phases \cite{tomography1}.\\
The TD system is solved with the help of the Lewis-Riesenfeld (LR) phase space invariant method. A toy model is explicitly solved analytically. The toy model is chosen in such a manner that the nonlinear integrable Ermakov-Pinney equation provides a solution in terms  of elementary functions. The invariant operator is solved after transforming it into a normal form with the help of symplectic diagonalization. Obtained coherent states then studied in detail. The full computation is done for the equivalent Hermitian partner for our target $\hat{\mathcal{P}}\hat{\mathcal{T}}$-symmetric oscillator corresponding to parametric amplification. The restrictions on the parameters for unbroken $\hat{\mathcal{P}}\hat{\mathcal{T}}$ symmetry region have been identified. Moreover, for general TD  parameters (in unbroken $\hat{\mathcal{P}}\hat{\mathcal{T}}$) we have obtained the exact solution for the system. TD phase-factors (both geometrical and dynamical) have been obtained for the solution of TD- Schr\"{o}dinger equation through the LR method. In general, the LR-invariant operator consists of a class of TD parameters. All the parameter values (coefficients of basic operators) might not be physically acceptable. For instance, we have employed the natural restriction of Robertson-Schr\"{o}dinger uncertainty principle (RSUP) to identify the domain of acceptable parameters.\\
We have studied the entanglement behavior only for two artificially prepared states, which is too much of a  simplification. For future study, the aspects of TD-amplification for arbitrary oscillators in $Sp(2n,\mathbb{R})$ will be an important study. Particularly, the separability criterion will be an interesting one.
\section{Data Availability Statement }
The manuscript has no associated data.
\section{Statements and Declarations}
{\bf Funding:}  No funding information to disclose.    \\ 
{\bf Conflict of Interest:} The author declare no competing interests.\\
{\bf Ethical Conduct:} Not applicable.
\section{Acknowledgement}
We are grateful to the anonymous referee for helpful suggestions. We are grateful to Abhi Mukherjee, University of Kalyani for his support in formatting the images.

\end{document}